\documentclass{iauJDSS}
\usepackage{graphicx}

\title[Solenoidal versus compressive turbulence forcing] 
{Solenoidal versus compressive turbulence forcing}

\author[Federrath et al.]   
{C.~Federrath$^1$, J.~Duval$^2$, R.~S.~Klessen$^1$, W.~Schmidt$^3$ \and M.-M.~Mac Low$^4$}

\affiliation{$^1$Zentrum f\"ur Astronomie der Universit\"at Heidelberg, 
Institut f\"ur Theoretische Astrophysik, \break
Albert-Ueberle-Str.~2, D-69120 Heidelberg, Germany\\[\affilskip]
$^2$Astronomy Department at Boston University, \break
725 Commonwealth Avenue, Boston, MA 02215, USA\\[\affilskip]
$^3$Institut f\"ur Astrophysik G\"ottingen, \break
Friedrich-Hund-Platz 1, 37077 G\"ottingen, Germany\\[\affilskip]
$^4$Department of Astrophysics, American Museum of Natural History, \break
Central Park West at 79th Street, New York, NY 10024-5192, USA
}

\pubyear{2009}
\volume{}  
\pagerange{??}
\date{?? and in revised form ??}
\jname{JD14}
\editors{M.~Cunningham}
\begin{document}

\maketitle

\begin{abstract}
We analyze the statistics and star formation rate obtained in high-resolution numerical 
experiments of forced supersonic turbulence, and compare with observations. We 
concentrate on a systematic comparison of solenoidal (divergence-free) and 
compressive (curl-free) forcing (Federrath et~al.~2009~a,b), which are two limiting cases of 
turbulence driving. Our results show that for the same RMS Mach number, 
compressive forcing produces a three times larger standard deviation of the density probability 
distribution. When self-gravity is included in the models, the star formation rate is more than 
one order of magnitude higher for compressive forcing than for solenoidal forcing.

\end{abstract}

Observational data indicate that turbulence in the ISM exhibits both signatures 
of solenoidal and compressive forcing, depending on the region under 
consideration (Heyer et~al.~2006;~Hily-Blant et~al.~2008;~Goodman et~al.~2009). In particular, 
expanding shells show statistical characteristics similar to compressively driven turbulence.

Compressive forcing produces a three times larger standard deviation 
of the three-dimensional and the column density probability distributions (PDFs) for the same RMS 
Mach number (Federrath et~al.~2008). When self-gravity is added to the models, the star 
formation rate is about 25 times higher for compressive forcing. These two results demonstrate 
that star formation models based on the turbulent density PDF 
(Padoan~\&~Nordlund~2002;~Krumholz~\&~McKee~2005;~Elmegreen~2008;~Hennebelle~\&~Chabrier~2009)
must take the nature of the turbulence forcing into account.


\end{document}